\documentstyle[12pt,epsf]{article}

\newcommand{\doublespace}{
  \renewcommand{\baselinestretch}{1.6}\large\normalsize}
 
\begin{document}

\hyphenation{plaq-uette}
\newcommand{\ts}{\textstyle}
\newcommand{\tr}{{\rm Tr}}

\begin{titlepage}

\begin{tabbing}
\` {\sl hep-lat/9612021} \\
    \\
\` LSUHE No. 249-1996 \\
\` HD-THEP-96-58 \\
\` ILL-(TH)-96-17 \\
\` OUTP-96-75P \\
\` December, 1996 \\
\end{tabbing}
 
\vspace*{1.0in}
 
\begin{center}
{\bf On the Phase Diagram of the SU(2) Adjoint Higgs Model
in 2+1 Dimensions\\}
\vspace*{.5in}
A. Hart$^1$, O. Philipsen$^2$, J. D. Stack$^3$ and M. Teper$^4$\\
\vspace*{.2in}
$^1${\it Department of Physics and Astronomy, Louisiana State University,\\
Baton Rouge. LA 70803 U.S.A.\\}
\vspace*{.2in}
$^2${\it Institut f\"ur Theoretische Physik, Universit\"at Heidelberg,\\
Philosophenweg 16, D-69120 Heidelberg, Germany.\\}
\vspace*{.2in}
$^3${\it Department of Physics,\\
University of Illinois at Urbana-Champaign,\\
1110 W. Green Street, Urbana. IL 61801 U.S.A. \\}
\vspace*{.2in}
$^4${\it Theoretical Physics, University of Oxford,\\
1 Keble Road, Oxford. OX1 3NP U.K.\\}
\vspace*{0.2in}
{\it (to be published in Physics Letters)}
\end{center}

\hspace{2.0in}
\end{titlepage}
\vfill\eject

\doublespace
\pagestyle{empty}

\begin{center}
{\bf Abstract}
\end{center}

	The phase diagram is investigated for $SU(2)$ lattice gauge
	theory in $d=3$, coupled to adjoint scalars.  For small values
	of the quartic scalar coupling, $\lambda$, the transition
	separating Higgs and confinement phases is found to be
	first-order, in agreement with earlier work by Nadkarni.  The
	surface of second-order transitions conjectured by Nadkarni,
	however, is shown instead to correspond to crossover
	behaviour.  This conclusion is based on a finite size analysis
	of the scalar mass and susceptibility. The nature
	of the phase transition at the termination of first-order
	behaviour is investigated and we find evidence for a critical
	point at which the scalar mass vanishes. The photon mass and
	confining string tension are measured and are found to be
	negligibly small in the Higgs phase.  This is correlated with
	the very small density of magnetic monopoles in the Higgs
	phase.  The string tension and photon mass rise rapidly as the
	crossover is traversed towards the symmetric phase.

\vspace*{.5in}
 
\setcounter{page}{0}
\newpage
\pagestyle{plain}

	In this paper we report on our investigation of the phase diagram
	for $d=3$ $SU(2)$ lattice gauge theory coupled to adjoint scalars.
	This theory is of interest for at least two reasons.  Through
	dimensional reduction, it is related to $d=4$ pure $SU(2)$ lattice
	gauge theory at high temperature.  The model also contains in its
	Higgs phase the famous 't Hooft-Polyakov monopoles.  The
	Higgs phase is continuously connected to the symmetric, confining 
	phase.  Thus the model may prove to be a useful laboratory for 
	exploring confinement by monopoles.  We will not further discuss
	monopole confinement here, but
	concentrate on the phase diagram and some general properties of the
	mass spectrum. 

	The continuum Euclidean action in the notation of Nadkarni 
\cite{nadkarni}
	is given
	by
	\begin{equation}
	S = \int d^{3}x \left\{ \frac{1}{2g^{2}} \tr(F_{\mu\nu}F_{\mu\nu})
	+\tr(D_{\mu}\phi D_{\mu} \phi) +m_{0}^{2} \tr(\phi \phi)
	+\frac{\mu}{2}(\tr(\phi \phi))^{2} \right\} ,
	\end{equation}
	where $F_{\mu\nu}=\partial_{\mu}A_{\nu}-\partial_{\nu}A_{\mu}
	+i[A_{\mu},A_{\nu}]$, 
	$D_{\mu}\phi = \partial_{\mu}\phi + i[A_{\mu},\phi]$; 
	and $F_{\mu\nu},A_{\mu}$, and $\phi$ are all traceless $2\times 2$ 
	Hermitian matrices ($\phi=\phi^{a}\sigma_{a}/2,etc)$.  
	The lattice form of the action is
	\begin{eqnarray}
	S &=& \beta \sum_{x,\mu > \nu}\left(1-\frac{1}{2}
	\tr U_{\mu\nu}(x)\right)
	+2\sum_{x} \tr(\Phi(x)\Phi(x)) \nonumber \\
	&-&2\kappa \sum_{x,\mu}\tr(\Phi(x)U_{\mu}
	(x)\Phi(x+\hat{\mu}a)U^{\dagger}(x))
	+\lambda \sum_{x}(2 \tr(\Phi(x)\Phi(x))-1)^{2}
	\label{lattice_action}
	\end{eqnarray}

	Here $U_{\mu}(x)=\exp(iagA_{\mu}(x))$ is the usual link variable,
	and $U_{\mu\nu}(x)$ is the corresponding plaquette.  The 
	continuum field $\phi(x)$ is related to the lattice field $\Phi(x)$ 
	by $\phi(x)=\sqrt{\kappa/a}\Phi(x)$, where $a$ is the lattice spacing,
	and
	the lattice parameters $\beta$ and $\lambda$ are defined by
	\begin{equation}
	\beta=\frac{4}{g^{2}a},\; \lambda=\frac{\mu a \kappa^{2}}{8}.
	\label{lattice_parameters}
	\end{equation}
	The hopping parameter $\kappa$ is determined by an equation relating
	it to the continuum mass parameter $m_{0}^{2}$.  The tree level form
	of this equation is
	$$
	m_{0}^{2} 
	= \frac{2}{a^{2}}\left[\frac{1}{\kappa} - 3 
                             -\frac{2\lambda}{\kappa}\right].
	$$
	Dividing by a factor $g^4$, and using Eq.(\ref{lattice_parameters}),
	this can be rewritten as
	\begin{equation}
	\frac{m_{0}^{2}}{g^4}= \frac{\beta^2}{8}\left[\frac{1}{\kappa}-3
	-\frac{\kappa}{\beta} \frac{\mu}{g^2}\right].
	\label{tree_constant_physics}
	\end{equation}
	Due to the superrenormalisability of the adjoint Higgs theory in
	$d=3$, $\lambda$ and $g^2$ do not require ultraviolet renormalisation.
	Suppose for the moment that this were true for $m_{0}^{2}$ as well.
	Eq.(\ref{tree_constant_physics}) then specifies how to adjust
	$\kappa$ to correspond to given fixed ratios of the continuum 
	parameters $m_{0}^{2}/g^{4}$ and $\mu/g^{2}$.   The 
	lattice spacing is set by $a=4/g^{2}\beta$.

	The mass parameter $m_{0}^{2}$ does of course require
	ultraviolet renormalisation, but only up to two loops.  The
	known results in the $\overline{\hbox{\sl MS}}$ scheme \cite{MS} and in
	lattice perturbation theory \cite{laine} allow a
	generalisation of Eq.(\ref{tree_constant_physics}) to be
	derived,
	\begin{eqnarray}
	\frac{m_{0}^{2}(g^{2})}{g^4}&=& \frac{\beta^2}{8}
                \left(\frac{1}{\kappa}-3
		-\frac{\kappa}{\beta} \frac{\mu}{g^2}\right) \nonumber \\
	        &+&\frac{\Sigma\beta}{4\pi}
                     \left(1+\frac{5\mu}{8g^{2}}\right) \nonumber\\
		&+&\frac{1}{16\pi^{2}}
                  \left[\frac{10\mu}{g^{2}}\left(1-\frac{\mu}{4g^{2}}\right)
	          \left(\ln\left(\frac{3\beta}{2}\right)+0.09\right)
                  +8.7+\frac{5.8\mu}{g^{2}}\right],
	\label{two_loop_constant_physics}		
	\end{eqnarray}
	where $\Sigma=3.17591$. The successive lines of this equation
	correspond to tree level, one loop and two loop terms,
	respectively.  The dimensionful parameter in the
	$\overline{\hbox{\sl MS}}$ scheme has been chosen to be
	$g^{2}$.
	Eq.(\ref{two_loop_constant_physics}) 
	specifies how to approach the
	continuum limit 
	($\beta \rightarrow \infty, \lambda \rightarrow 0, 
	\kappa \rightarrow 1/3$),
	with $m_{0}^{2}(g^{2})/g^{4}$ and $\mu/g^{2}$ 
	held fixed. It
        becomes exact in the limit of vanishing lattice spacing.

	Our analysis of the phase diagram is based on the early work
	of Nadkarni \cite{nadkarni}, who first investigated this model
	using a combination of numerical and analytic techniques.
	Following Nadkarni, we define the length $\rho$ of the scalar
	field by $ \rho = \sqrt{2 \tr(\Phi\Phi)} $.  Qualitatively the
	Higgs phase and symmetric phases can be distinguished by the
	value of $\langle \rho \rangle$, it being larger in the Higgs
	phase.  The familiar language of symmetry breaking is useful
	in the Higgs phase, whereas the symmetric phase is better
	described as a confined hadronic phase.  As was emphasized by
	Nadkarni, however, the two phases are in fact continuously
	connected, and there is really no fundamental distinction
	between them.
	When an actual phase transition separates the two phases, it
	is described by a surface $\kappa_{c}(\beta, \lambda)$, the
	Higgs (symmetric) phase lying above (below) the surface.  The
	statement that the two phases are continuously connected
	implies that this surface does not cover the entire
	$\beta-\lambda$ plane, so that any two points in the parameter
	space can be connected by a path which detours around it.
	While it is not feasible to map out the entire surface in a
	strictly numerical analysis, one of our principal results will
	be that the surface of phase transitions covers a much smaller
	region of the $\beta-\lambda$ plane than conjectured by
	Nadkarni.  As discussed in more detail below, the vast
	region of the surface specified by Nadkarni as second-order
	corresponds rather to a crossover region which is without
	critical fluctuations or massless particles.

	Our methods are similar to those used by two of the present
	authors in a previous study in $d=3$ with $SU(2)$ gauge fields
	coupled to fundamental rather than adjoint scalars
	\cite{fundamental}.  
        The link update used in
	\cite{fundamental} was modified by inserting a Metropolis step
        to take account of the
	quadratic dependence of the scalar term in
	Eq.(\ref{lattice_action}) on the link variables.  The Bunk
	\cite{bunk} algorithm for the scalar update was used here also.
	With adjoint scalars, however, the
	approximate determination of Bunk's parameter $\alpha$ suggested in
        \cite{bunk} and used in \cite{fundamental}
	occasionally resulted in a low acceptance 
	rate for the scalar update (particularly at very low $\beta$).  
	Solving the full
	cubic equation (Eq.(6) of \cite{bunk}) for $\alpha$ gave
	acceptance rates of over 80\% 
	throughout the phase diagram.

	The results presented here were obtained on lattices of size
	$L^{3}$ with $L=8$, 12, 16, 20, 24 and 32 at $\beta=6.0$ and
	$9.0$.  A typical run consisted of 500 -- 2000 sweeps.  The coupling
	$\beta = 6.0$ is already in the region of approximate
	scaling for $d=3$ pure $SU(2)$ lattice gauge theory
	\cite{teper_3}. In the case with fundamental scalars
	\cite{fundamental}, the mass spectrum at the nearby value of 
	$\beta=7.0$ has been calculated and has been found to
	scale very well to higher values of $\beta$.
	In the present case of adjoint scalars, we shall see later
	that there is good scaling of mass ratios between $\beta=6.0$
	and 9.0. The procedure is, given $\kappa$ and $\lambda$ at
	say, $\beta=6.0$, to use Eqs.(\ref{lattice_parameters}) and
	(\ref{two_loop_constant_physics}) to determine the ratios of
	continuum parameters $m_{0}^{2}(g^{2})/g^{4}$ and $\mu/g^{2}$.
	Then holding these ratios fixed,
	Eqs.(\ref{lattice_parameters}) and
	(\ref{two_loop_constant_physics}) are solved again to find 
	$\kappa$ and $\lambda$ at the new value of $\beta$,
	e.g. $\beta=9.0$.  Close enough to the continuum limit,
	physical mass ratios should be constant when moving along such
	a `line of constant physics'.

        \begin{figure}[th]
        \begin{center}
        \leavevmode
        \epsfysize=250pt
        \epsfbox[20 30 620 730]{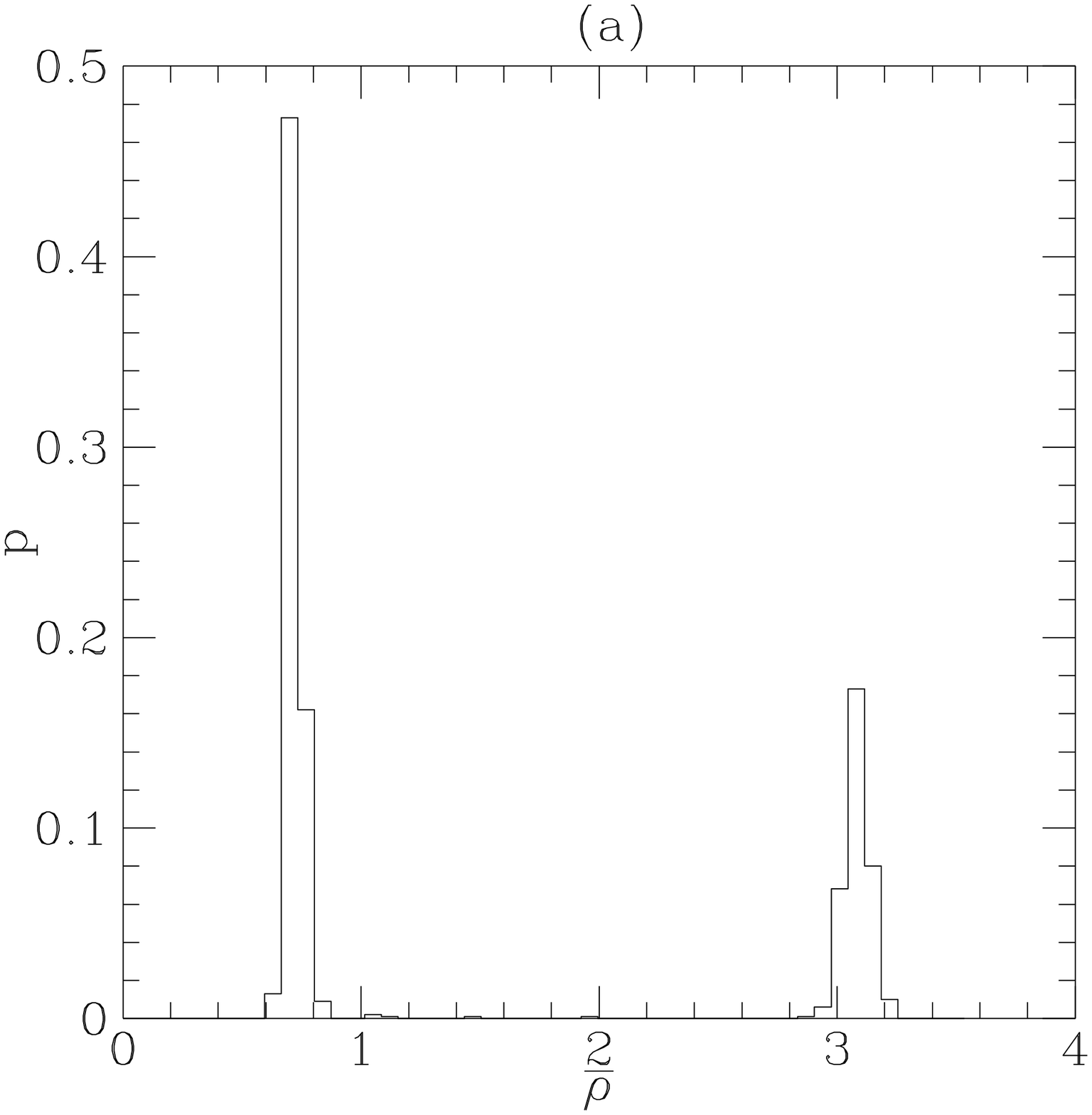}
        \leavevmode
        \epsfysize=250pt
        \epsfbox[20 30 620 730]{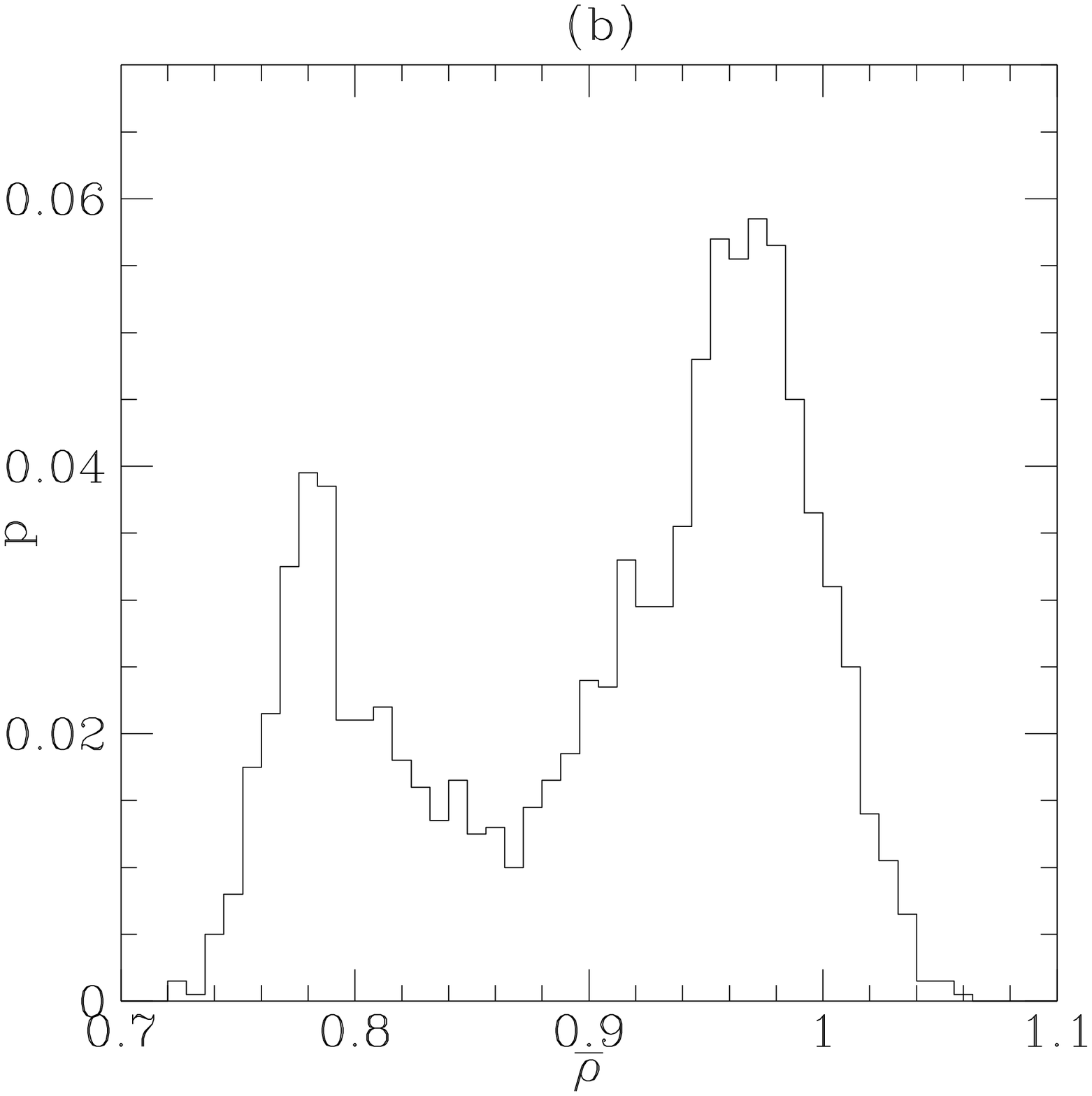}
        \vspace{-1.6cm}
        \end{center}
        \caption[]{\it \label{hist}
                   Probability distributions of $\bar{\rho}$
                   indicating a first-order phase transition
                   at $\beta=6.0$. (a) $\kappa=0.370473,\lambda=0.001, L=8$.
                   (b) $\kappa=0.38887, \lambda=0.007, L=20$.} 
        \end{figure}

	The most straightforward region of the phase diagram to
	analyse is where Higgs and symmetric phases are separated by a
	first-order transition.  In agreement with Nadkarni, we find
	clear first-order behaviour for small $\lambda$, which for
	$\beta=6.0$ means $\lambda \ll 0.01$.  In Fig.~\ref{hist}(a)
	we show the histogram of $\bar{\rho}$ at $\beta=6.0,
	\lambda=0.001$ on an $8^{3}$ lattice. $\bar{\rho}$ is the
	average over the lattice volume, $V$, of $\rho(x)$. Two peaks
	corresponding to symmetric and Higgs phases are clearly seen.
	Tunnelling between the two states occurred once in this
	run. When we increased the lattice size the peaks became so
	distinct that the to-and-fro tunnelling became unobservably
	rare.  The first-order nature of the transition can also be
	seen from a plot of $\langle \bar{\rho} \rangle$ versus $\kappa$.
	There is a jump from smaller to larger values of $\langle \bar{\rho}
	\rangle$ which becomes rapidly sharper as the lattice volume
	is increased.  The first-order behaviour observed at $\lambda
	= 0.001$ persists but weakens as $\lambda$ increases. The two
	peaks in $\bar{\rho}$ become less widely separated and
	tunnelling between the two states becomes more frequent, {\it
	e.g.}  Fig.~\ref{hist}(b) for $\lambda = 0.007$. Here the
	barrier is weak and we have frequent tunnelling, which
	persists even on fairly large lattices. Clearly at this value
	of $\lambda$ the first-order transition has become very weak
	and, as we shall see below, for slightly larger $\lambda$ one
	loses all evidence for first-order behaviour.  We see exactly
	the same behaviour in the corresponding calculations at
	$\beta=9.0$.

        \begin{figure}[th]
        \begin{center}
        \leavevmode
        \epsfysize=180pt
        \epsfbox[20 30 620 730]{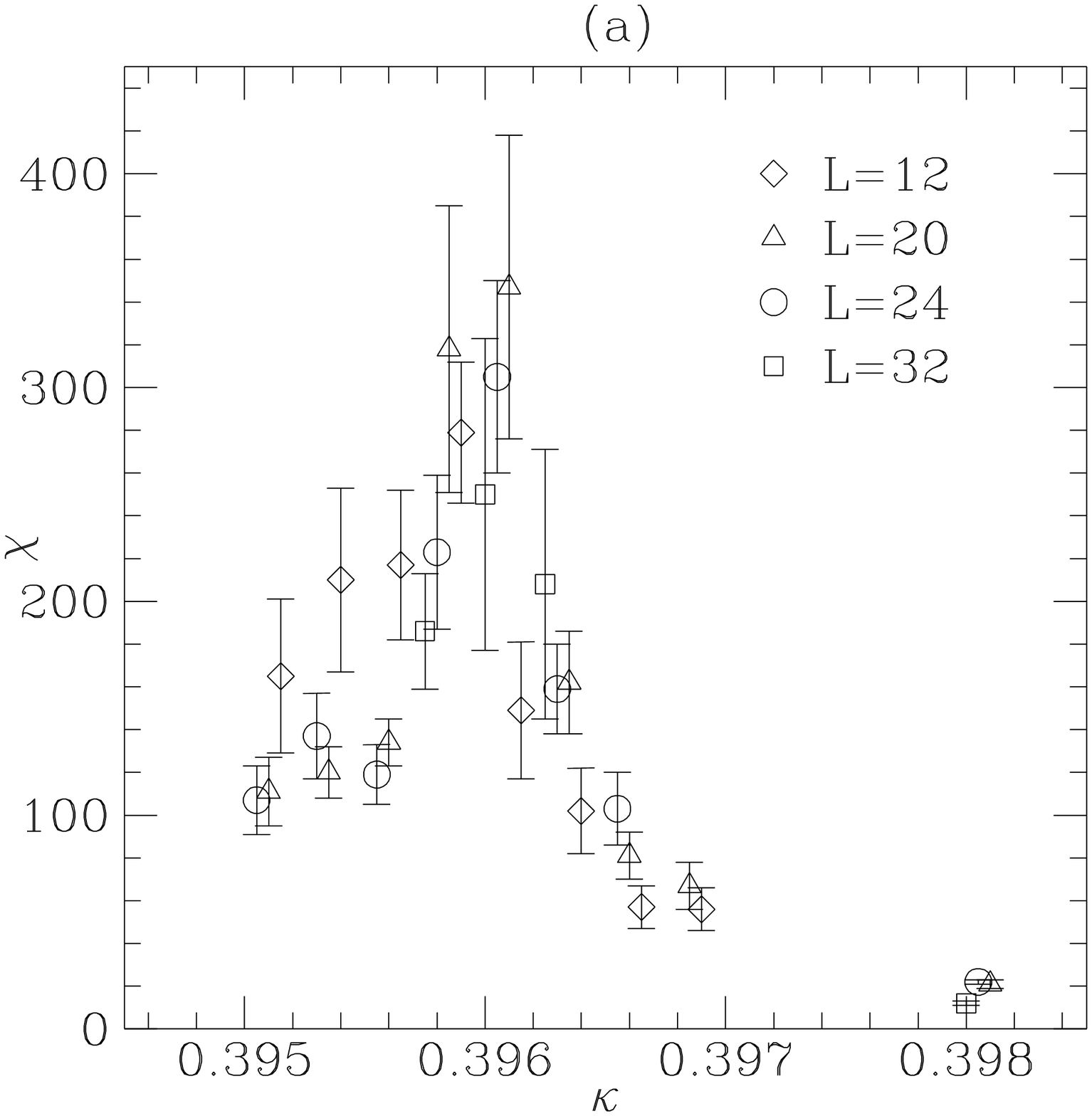}
        \leavevmode
        \epsfysize=180pt
        \epsfbox[20 30 620 730]{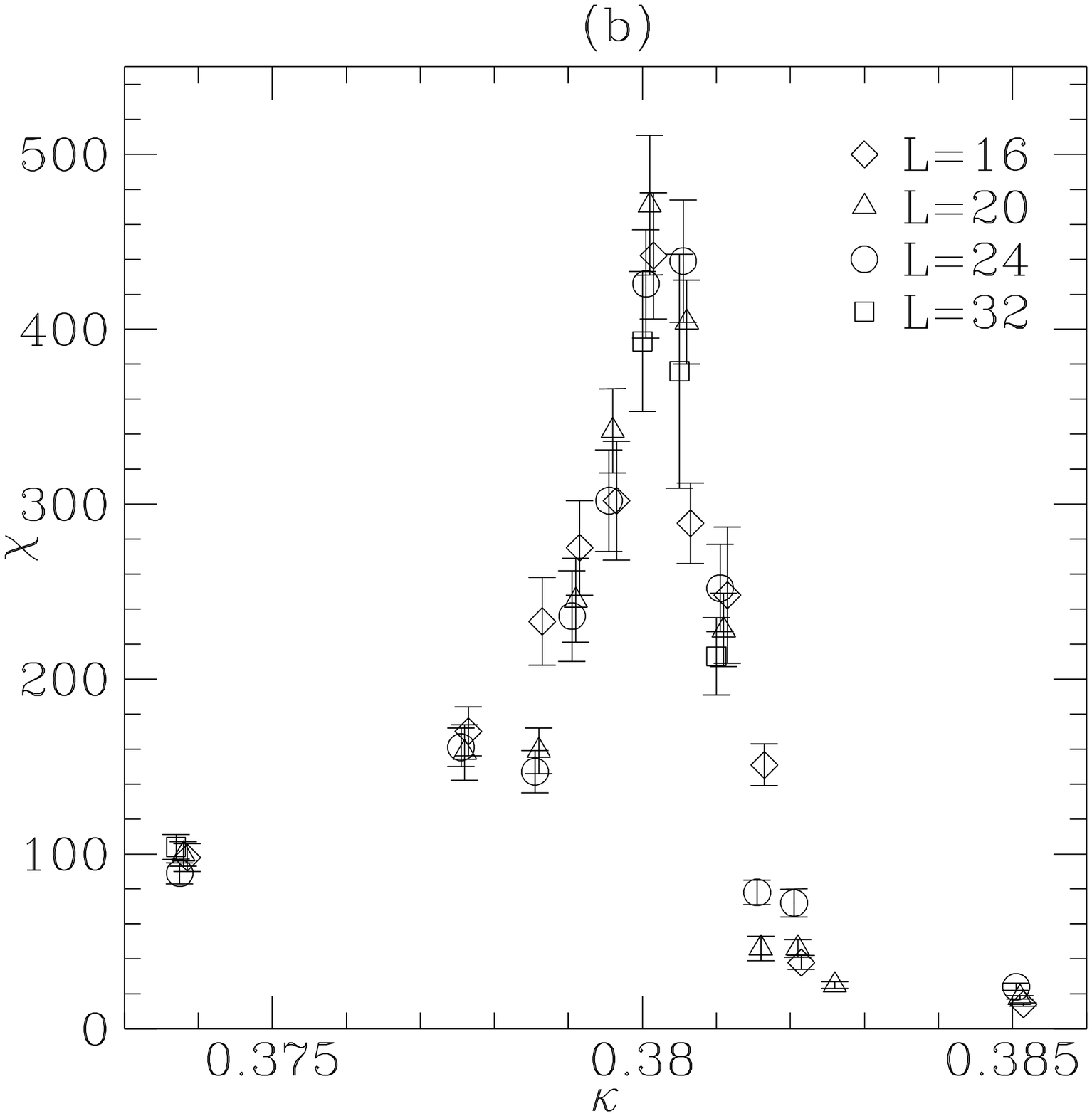}
        \leavevmode
        \epsfysize=180pt
        \epsfbox[20 30 620 730]{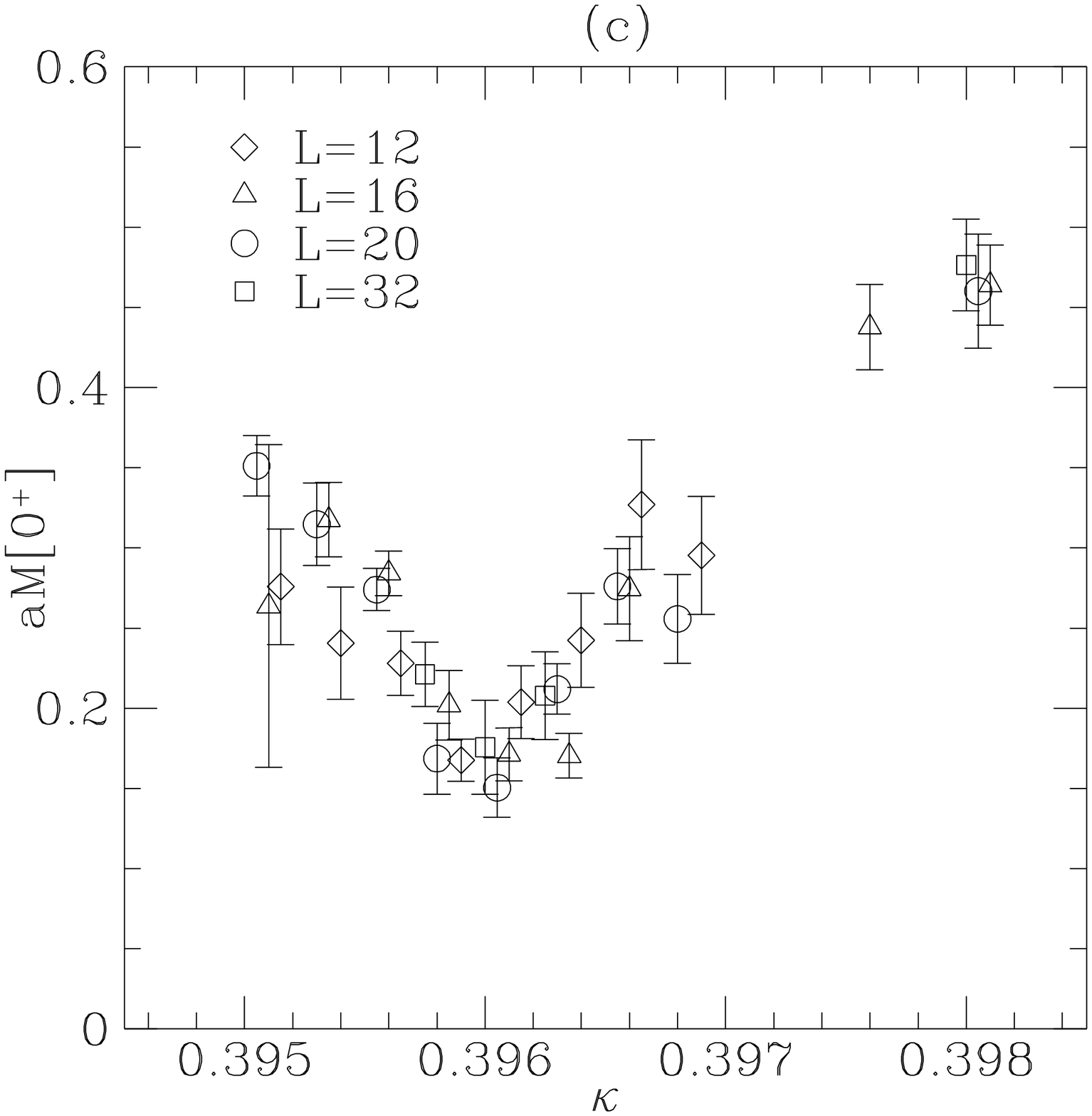}
        \vspace{-1.6cm}
        \end{center}
        \caption[]{\it \label{cross}
                   Finite size scaling of scalar susceptibility and mass
                   indicating crossover behaviour.
                   	(a) $\beta=6.0, \lambda=0.01$. 
			(b) $\beta=9.0, \lambda=0.01$.
                   	(c) $\beta=6.0, \lambda=0.01$.}
        \end{figure}

	Let us defer for the moment the question of what happens at
	the precise value of $\lambda$ (for fixed $\beta$) at which
	first-order behaviour terminates.  It was conjectured by
	Nadkarni that beyond the region of first-order behaviour, the
	surface $\kappa_{c}(\beta,\lambda)$ continues (all the way to
	$\lambda = \infty$ at $\beta=6.0$ and $9.0$), describing a
	second-order transition. We instead find no actual phase
	transition, only a smooth crossover behaviour.  This conclusion
	is reached using a finite size analysis of the scalar mass and
	related susceptibility.
	Our scalar
	mass calculations are based on the operator
	$\rho_s(x) \equiv \rho^2(x)$ (and smeared versions thereof).
	The corresponding (normalised) susceptibility is
	$$
	\chi = \frac{V}{a^3} {\biggl (}
	{{\langle \bar{\rho_s}^2 \rangle}\over{\langle \bar{\rho_s}\rangle^2}}
	-1 {\biggr )},
	$$
	where $V$ is the lattice volume. Fig.~\ref{cross}(a) shows how
	this susceptibility varies with $\kappa$ at $\beta=6.0$ and
	$\lambda=0.01$. While there is a pronounced peak, it neither
	grows nor sharpens as $V$ is increased (except perhaps for
	very small volumes). Fig.~\ref{cross}(b) shows a similar plot,
	with similar behaviour, taken at $\beta=9.0$.  In
	Fig.~\ref{cross}(c) we show how the lightest scalar mass
	varies across this peak. (This is calculated from smeared
	versions of $\rho_s$; the smearing is exactly as in
	\cite{fundamental} except that fundamental parallel
	transporters are replaced by their adjoint homologues.)
	According to a general finite size scaling analysis
	\cite{barb} a true second-order transition would be signalled
	by a peak in the susceptibility growing as $\chi_{max}\sim
	V^\gamma$ with some critical exponent $\gamma$.  Likewise,
	across a second-order transition the lightest mass should go
	to zero with increasing lattice size, corresponding to a
	diverging correlation length. In the vicinity of the peak of
	the susceptibility we found the lightest scalar to be the
	lightest excitation. Although we do observe a susceptibility
	peak along with a minimum in the scalar mass, neither shows
	the required finite-size behaviour for a second-order
	transition.  Instead we find $\chi,aM[0^+]\rightarrow
	\hbox{const}$ for $V\rightarrow \infty$, which is the expected
	behaviour for crossover.  The same observation holds at still
	larger values of $\lambda$ and at $\beta=9.0$ as well.

	Where the first-order transition ends and the crossover
	begins, one would expect to find critical behavior, $i.e.$
	diverging fluctuations and a vanishing mass for the lightest
	scalar.  This would form a line of critical points in our
	three dimensional coupling constant space.  To investigate
	this, we calculate the minimum value of the mass of the scalar
	as we traverse the crossover, and then move along the line of
	minima towards the first-order region.  Suppose we work at
	fixed $\lambda$ and $\beta$, and find the minimum scalar mass
	$m_{min}(\beta,\lambda)$ at
	$\kappa=\kappa_{min}(\beta,\lambda)$.  If we now steadily
	decrease $\lambda$ at fixed $\beta$, a critical point
	$\lambda_c(\beta)$, may occur at the boundary between
	crossover and first-order regions.  This would be signalled by
	a vanishing of the scalar mass: $m_{min} \propto
	(\lambda-\lambda_c)^\nu$.  We have performed such a study at
	$\beta=6.0$ in which we have calculated $m_{min}$ for three
	values of $\lambda$. Since $m_{min}$ is small, it is crucial
	that we ensure that the lattices we use are large enough. Our
	scans through the crossover were carried out on a range of
	lattice sizes from $8^3$ to $32^3$.

	The results of these calculations are shown in Table \ref{end}.
        \begin{table}[ht]
        \begin{center}
        \begin{tabular}
{r@{.}l *{2}{r@{.}l@{\hspace{0.5\tabcolsep}(}r@{)\hspace{2\tabcolsep}}}}
        \hline \hline
        \multicolumn{2}{c}{$\lambda$} 
        & \multicolumn{3}{c}{$\kappa_{min}$}
        & \multicolumn{3}{c}{$a m_{min}$} \\ 
	\hline
        0&011 & 0&39820 & 30 & 0&207 & 23 \\ 
        0&010 & 0&39600 & 25 & 0&161 & 14 \\ 
        0&009 & 0&39380 &  5 & 0&116 &  8  \\
        \hline \hline
        \end{tabular}
        \end{center}
        \caption[]{\label{end} \it Minimal scalar mass in the crossover region
                    approaching $\lambda_c$ at $\beta = 6.0$.}
        \end{table}
	We observe that $m_{min}(\lambda)$ does decrease by
	about a factor of 2 when  $ \lambda$ is reduced from 0.011 to
	0.009. This is suggestive  of a critical point at a value
	of $\lambda < 0.009$.
	Recall that we have performed calculations at $\beta=6.0$ and
	$\lambda=0.007$
	which reveal a very weak first-order transition. This 
	would place the suspected critical point for $\beta=6.0$ in the
	interval $0.007 < \lambda_c(6.0) < 0.009$.
	To locate it with more precision would require long runs on large
	lattices.

	We consider now gauge bosons and magnetic
	monopoles.  Where the scalar particles are in the fundamental
	representation of $SU(2)$, the existence of a residual or
	custodial $SU(2)$ global symmetry allows operators which
	couple directly to the triplet of gauge bosons to be
	constructed and the gauge boson masses determined from
	correlators of these operators \cite{fundamental}.  For the
	present case of adjoint scalars, no such residual symmetry
	exists, and consequently there are no gauge-invariant
	operators which couple directly to charged gauge bosons.  A
	gauge-invariant
	operator which directly couples to the photon in the Higgs
	phase  can be constructed, however, starting with the
	continuum operator $2\tr(\phi F_{\mu\nu})$.  In the Higgs
	phase, $A^{3}_{\mu}$ is the electromagnetic gauge field in
	unitary gauge ($\phi = \phi^{3}\sigma_{3}/2$).  In this gauge
	our operator is $\sim F^{3}_{\mu\nu}$, and the latter is $\sim
	\partial_{\mu}A^{3}_{\nu}-\partial_{\nu}A^{3}_{\mu}$, since
	the gauge fields $A^{1,2}_{\mu}$ have only short range
	fluctuations.  Thus in the Higgs phase, $2\tr(\phi
	F_{\mu\nu})$ couples to the field strength
	of the photon, and we can determine the photon mass by
	measuring its correlator \cite{polyakov}.  Working as we do in
	momentum space, at least one of the momentum components
	$p_{\mu},p_{\nu}$ must be non-vanishing in order to retain
	that part of $2\tr(\phi F_{\mu\nu})$ which is $\sim
	\partial_{\mu}A^{3}_{\nu}-\partial_{\nu}A^{3}_{\mu}$.

	On the lattice, we start from 
	${\cal F}_{\mu\nu} = 2\tr(\Phi(x)U_{\mu\nu}(x))$
        which contains $2\tr(\phi F_{\mu\nu})$ in the continuum limit.
	The operator we use in practice has
	the field $\Phi$ inserted at all corners of 
	the plaquette to maintain symmetry,  and is then smeared
	to improve overlaps.
	We have measured
	the energy of the lowest state coupled to this operator
	at finite momentum.  

        \begin{figure}[t]
        \begin{center}
        \leavevmode
        \epsfysize=250pt
        \epsfbox[20 30 620 730]{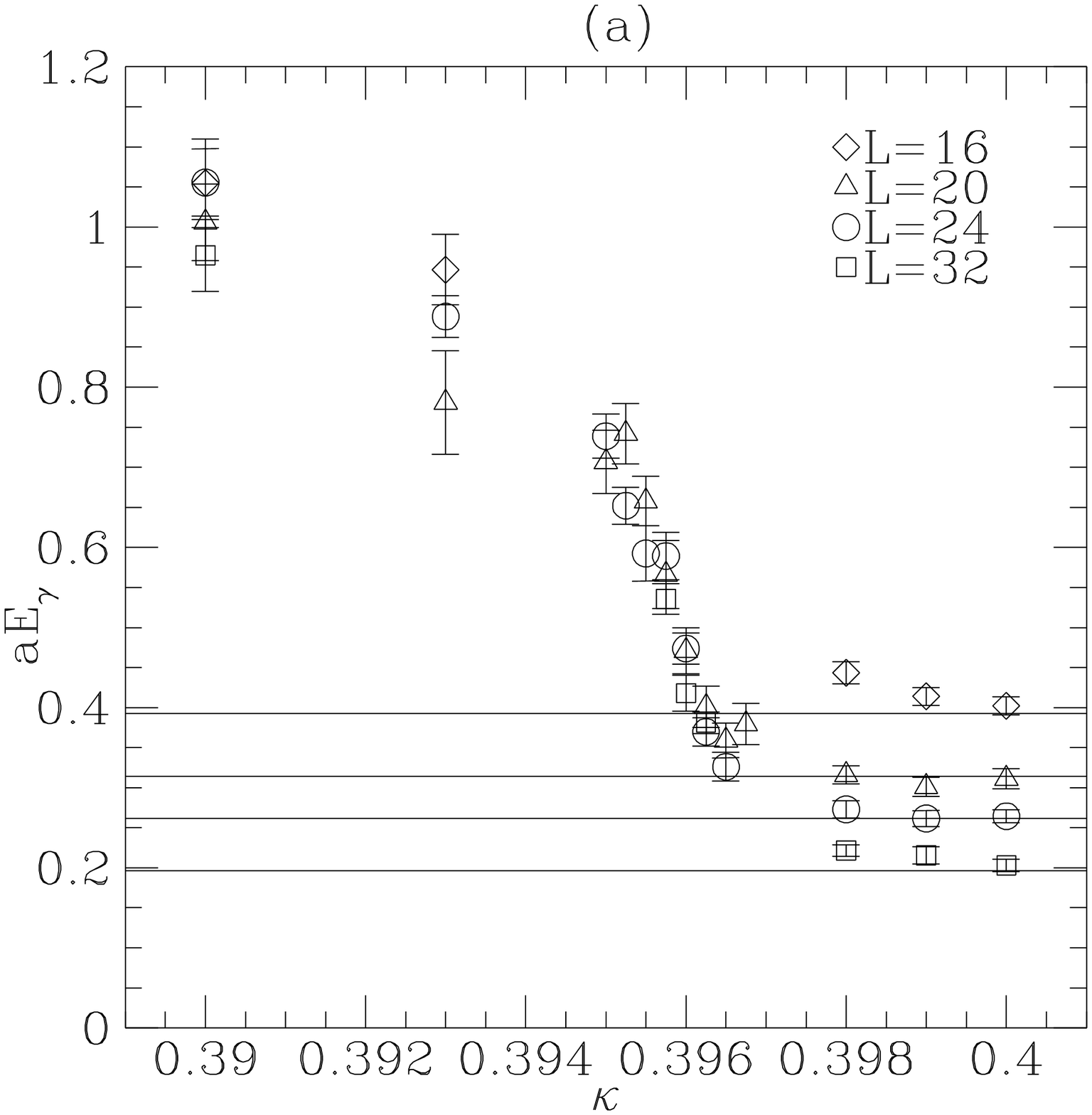}
        \leavevmode
        \epsfysize=250pt
        \epsfbox[20 30 620 730]{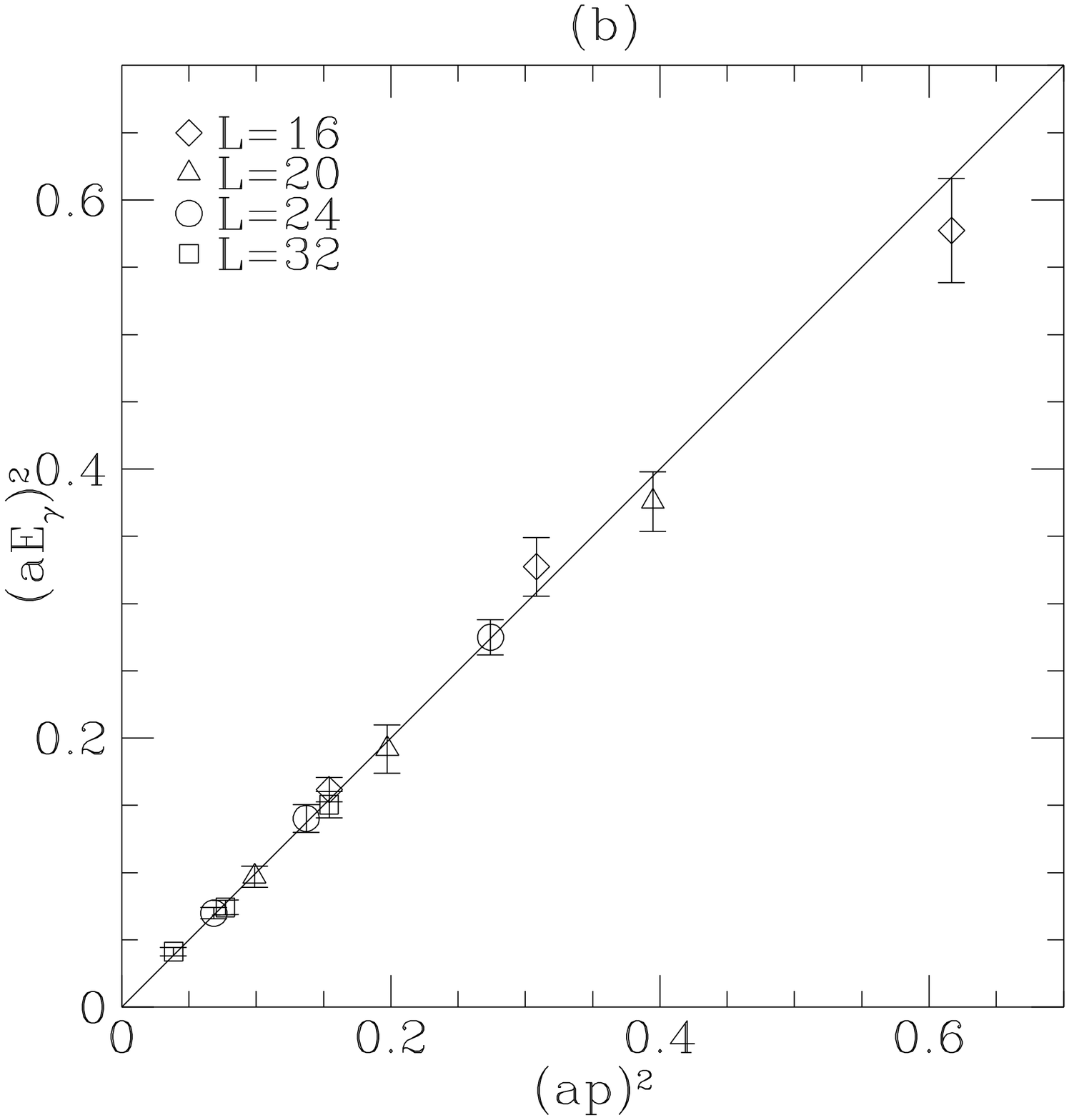}
        \vspace{-1.6cm}
        \end{center}
        \caption[]{\it \label{photon}
                   (a) Photon energy for $p=2\pi/L$ at $\beta=6.0, 
\lambda=0.01$.
                       The solid lines mark $p=2\pi/L$ for the respective
                       lattice sizes.
                   (b) Dispersion relation for the photon at
                       $\beta=6.0, \lambda=0.01, \kappa=0.4$.
                       The solid line marks the expectation for a 
                       massless photon.}
        \end{figure}
        Fig.~\ref{photon}(a) shows the 
	energy, $E(p)$, for the lowest non-zero lattice momentum,
	$p_{\mu}={{2\pi}\over L}(1,0)$ taken through the crossover at 
	$\beta=6.0$, $\lambda=0.01$.  
	As can be seen, once the value of $\kappa$ is
	in the Higgs phase, {\it i.e.} $\kappa >
	\kappa_{c}(6.0, 0.01)\sim 0.396$, the energies agree very well
	with $2\pi/L$ which is the energy of a massless excitation on
	a lattice of side $L$.  
	The results are equally good for all of the three lowest non-zero
	lattice momenta that we studied.  To demonstrate this we plot
	in Fig.~\ref{photon}(b) $E^2(p)$ versus $p^2$ for all $L$ and
	for the three lattice momenta at each $L$ at a point well inside the
	Higgs phase.  We see very striking evidence for the
	relativistic dispersion relation of a massless photon;
	a fit to the data yields 
	$(aM_{\gamma})^2 = 0.0010 (20)$.
        In contrast, as we decrease
	$\kappa$ and move into the symmetric phase, the energy rises rapidly as
	it should.  In the symmetric phase the operator in question
        describes an object composed of adjoint scalars
	and gauge particles or `gluons' which will have a mass on the
	hadronic scale. Equivalent results are obtained 
	at $\beta=9.0$ and other values of $\lambda$.

\begin{figure}[th]
        \begin{center}
        \leavevmode
        \epsfysize=250pt
        \epsfbox[20 30 620 730]{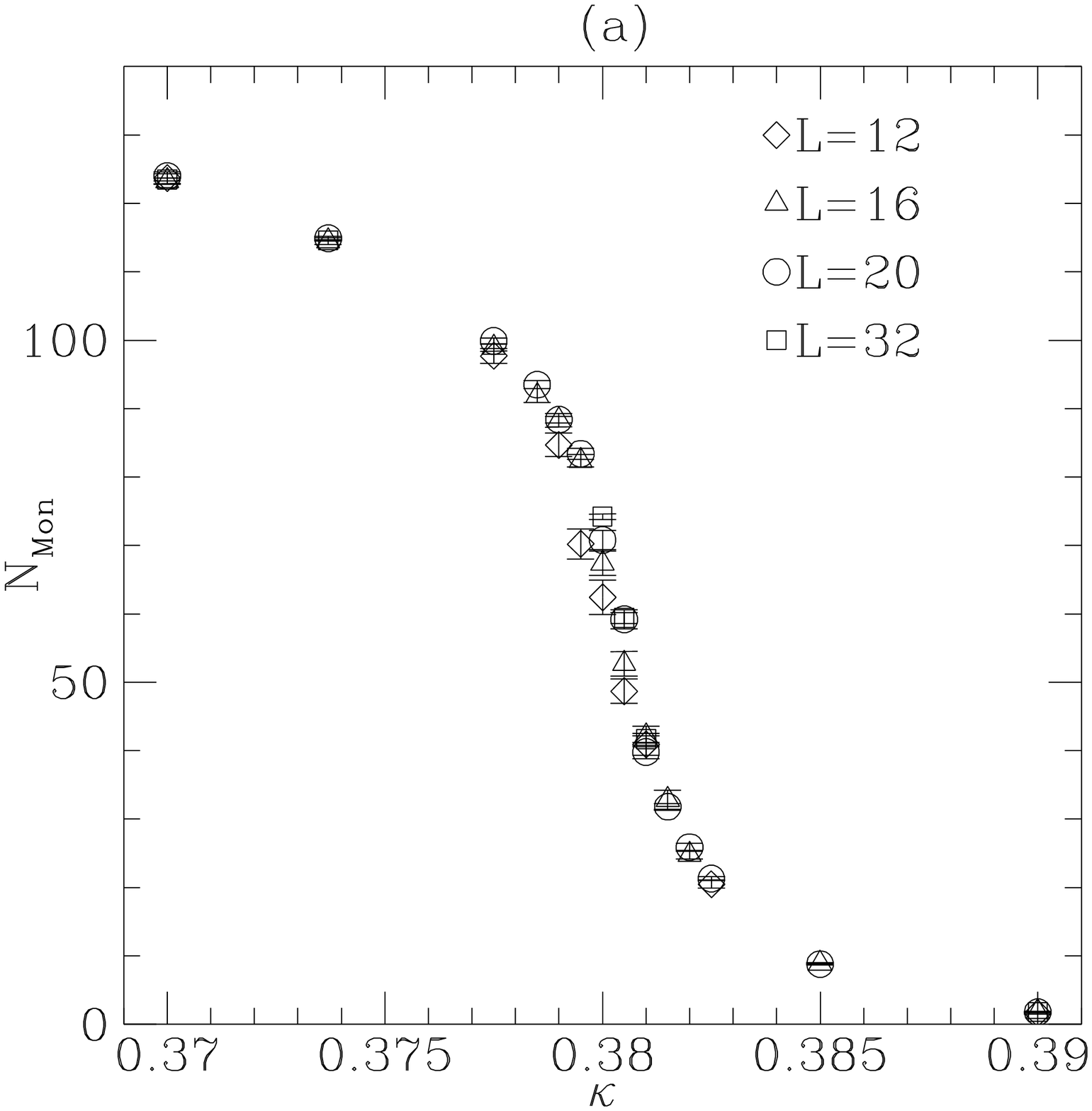}
        \leavevmode
        \epsfysize=250pt
        \epsfbox[20 30 620 730]{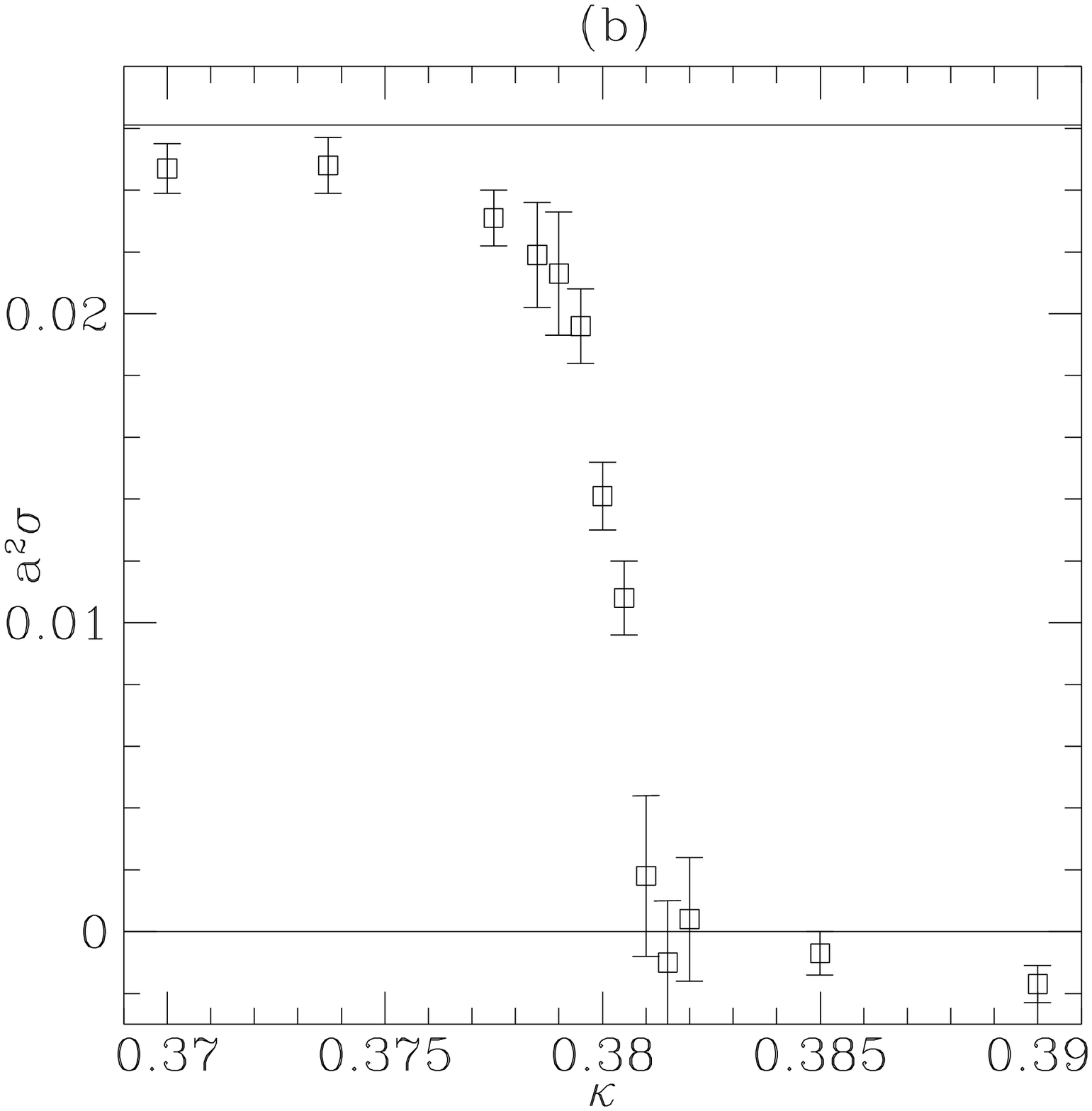}
        \vspace{-1.6cm}
        \end{center}
        \caption[]{\it \label{string}
          Properties of the crossover region at $\beta = 9.0, \lambda = 0.01$.
         (a) Number of monopoles in a $12^3$ cube.
         (b) String tension; 
             the upper solid line marks that in the pure gauge theory.}
        \end{figure}

	A photon of negligible mass in the Higgs phase is at first
	sight surprising.  The classic work of Polyakov \cite
	{polyakov} established for adjoint scalars coupled to $SU(2)$
	gauge fields, that the photon has a finite mass due to the
	presence of 't Hooft-Polyakov monopoles.  This apparent
	paradox is resolved by noting that for the present range of
	parameters, monopoles have a very large action and the photon
	mass established by Polyakov is far too small to show up on
	the lattices studied here.  
	An estimate for the action of a monopole is
	$S_{m} \sim 4\pi M_{W}/g^2$,where
	$M_{W}$ is the mass of the charged vector
	boson in the Higgs phase \cite{thooft}.
	We have not directly measured $M_{W}$ as this would 
        require gauge fixing (see the discussion above).
	We proceed by using 
        the tree level expression $M_W^2/g^4=-2m_0^2/(\mu g^2)$.
        Converting our parameter values $\beta=6.0, \lambda=0.01,
        \kappa \simeq 0.4$ to the continuum using 
        Eq.(\ref{two_loop_constant_physics}) we find $S_{m}\approx 5.8$.
	Now Polyakov's formulas  
	for the photon mass $M_\gamma$  and monopole
	density $n$ are \cite{polyakov},
	$$
	M_{\gamma}^{2} \sim \frac{2}{g^2}n,\; n 
	\sim \frac{(M_{W})^{\frac{7}{2}}}{g}\exp(-S_{m}).$$
	Evaluating these at $\beta=6.0$ gives $(M_{\gamma}a)^{2} \sim 10^{-4}$
	and $na^{3} \sim 10^{-4}$, so our small value for the photon mass
	is to be expected.
	We have measured the monopole density (in unitary gauge) using
	standard methods \cite{degrand} in 
	our configurations and find that there are indeed
	very few monopoles in the Higgs phase, 
        as is displayed in Fig.~\ref{string}(a).
        Of those that are there, all (within errors) are members 
	of tightly bound dipoles, and therefore ineffective in
	generating a photon mass. The signal for dipoles is that 
        the average flux from a monopole 
        vanishes into noise within one or two lattice 
        spacings away from the monopole location.

	If, for all practical purposes, the Higgs phase does not contain
	any magnetic monopoles then as we cross
	over from the symmetric phase to the Higgs phase, the
	string tension (the coefficient of the linear piece of the 
	confining potential) should vanish: $\sigma \to 0$. To
	investigate this we have calculated the mass of the
	lightest flux loop that joins onto itself through the 
	periodic boundary of the lattice. If we have linear
	confinement then the mass, $m_l$, of such a flux loop should
	satisfy
        \begin{equation} \label{mpol}
	a m_l(L) = a^2 \sigma L - {\pi \over {6L}} ,
        \end{equation}
	where we have included the well-known universal leading
	correction to the linear piece. We calculate $am_l(L)$
	from the asymptotic exponential decay of correlations of 
	smeared Polyakov loops on $L^3$ lattices exactly as 
	in the pure gauge theory. In addition to the standard
	Polyakov loops, $l_P$, which we may schematically define
	by $l_P = \tr \{\prod^L_{t=1} U_t\}$, we also consider modified
	Polyakov loops obtained by inserting scalar fields;
	{\it i.e.} $l^s_P = \tr\{ \Phi(L) \prod^L_{t=1} U_t\}$ and
	$l^{ss}_P = \tr\{\prod^L_{t=1} U_t\Phi(t)\}$. These operators
	have the same properties under the centre of the gauge group
	as the Polyakov loop and therefore project only onto
	states with non-trivial winding, such as our periodic flux loop.
	We mention this technical detail because, as we move through 
	the crossover region from the symmetric to the  Higgs phase,
	the best operators become $l^s_P$ and then $l^{ss}_P$ and
	one needs to use them rather than $l_P$ in order to
	obtain accurate mass estimates. (Of course, this observation
	tells us something interesting about the dynamics of the 
	crossover as well.)

	We show in Fig.~\ref{string}(b) a plot of our calculated value of
	the string tension, $a^2\sigma$, as we vary $\kappa$
	through the crossover at $\beta=9.0$ and $\lambda=0.01$.  
	The calculation was performed as follows. We calculated
	$m_l(L)$ for $L=16,20,24$ and, for perhaps half the $\kappa$-values,
	for $L=32$ as well. We then fitted these masses to the form
	Eq.~(\ref{mpol}). We obtained acceptable values of $\chi^2$
	for $\kappa \leq 0.3805$ and so for this range of $\kappa$
	this is how we obtained $\sigma$. (It is interesting to note
	that, throughout this range of $\kappa$, fits without the 
	$O(1/L)$ string correction are very much poorer, even in 
	the transition region where $a^2\sigma$ is rapidly decreasing.
	This provides strong statistical evidence for the presence
	of such a correction and, indeed, for its `universality'.)
	For $\kappa \geq 0.3810$,
	the value of $\chi^2$ increased sharply. The
	confidence level of the best fit dropped from $90\%$ at
	$\kappa=0.3805$ to  $0.5\%$ at $\kappa=0.3810$ and to 
	$<0.01\%$ at $\kappa \geq 0.3815$. In this range of
	$\kappa$ we therefore allowed a constant piece and 
	fitted the loop masses to $am_l(L) = c + a^2 \sigma L$.
	Such fits had an acceptable $\chi^2$.
	(One can include a string correction as above, but it makes
	no significant difference.) 

        We see in Fig.~\ref{string}(b) that the string tension collapses to
	zero around $\kappa \simeq 0.3805$. This is precisely
	the centre of the crossover ({\it i.e.} the coupling where the
	scalar mass attains its minimum value as we traverse
	the crossover at fixed $\beta$ and $\lambda$). For
	$\kappa \geq 0.3810$, $m_l(L)$ is constant to a first
	approximation (apparently with a correction that seems to 
	lead to a slight decrease with increasing $L$ so that
	the fitted $\sigma$ has a tendency to be slightly negative). 
	This is in fact what we would expect in a Higgs phase
	with no monopoles. The flux loop consists of a unit flux 
	traversing the $L\times L$ spatial plane in, say, the 
	$y$-direction. In the Higgs phase we would expect
	there to be no flux `tube' so that the flux simply 
	spreads out uniformly in the $x$-direction. Moreover,
	since we are in the $U(1)$-like Higgs phase, we neglect
	non-Abelian effects. 
        Then the field strength is
	$E \sim {1\over L}$, the energy density $\sim {1\over L^2}$
	and hence when we integrate this over the $L\times L$ plane
	we get a  total energy  that is independent of $L$. This
	matches what we observe and suggests that what happens to
	the confining flux tube as we traverse the centre
	of the crossover is that its width goes from being
	$O({1\over ag^2})$ to being $O({1\over L})$. This change
	takes place over a remarkably narrow range of couplings:
	between $\kappa=0.379$ and $\kappa=0.381$ in the case
	being considered here. Of course if we let $L \to \infty$
	then there will be a very dilute gas of monopoles
	and this will reintroduce a finite, albeit extremely wide, 
	flux tube and a non-zero, albeit extremely small, $\sigma$.
	Finally we remark that although the Higgs phase may have
	no visible string tension, it is still confining, albeit
	only logarithmically, because we are in two spatial
	dimensions. Thus $\langle l_P \rangle = 0$ for
	all values of the coupling and there is no $need$ for a phase
	transition. 
	
	We return briefly to the lines of constant physics.  At
	$\beta = 9.0$, $\kappa = 0.3737$, $\lambda = 0.01$ we are deep
	in the symmetric phase. Using a variety of operators we
	measure masses and the string tension as above. Now
	along a  line of constant physics the ratio of a
	mass to the square root of the string tension should be
	independent of the lattice spacing scale, if the $O(a)$
	corrections are small. To test this, we use
	Eqs.(\ref{lattice_parameters}) and (\ref{two_loop_constant_physics})
	to map $\kappa = 0.3737$, $\lambda = 0.01$ to the
	corresponding values at $\beta = 6.0$, and see whether
	the two sets of calculated mass ratios agree.
	This is indeed what we find for operators of purely gluonic,
	purely scalar or mixed origin (within errors 
	which are $< 10\%$). Repeating this for a 
	$\beta=9.0$ point in the crossover region 
	($\lambda=0.01,\kappa = 0.3805$),
	agreement is still good, although less so for the admittedly
	more noisy gluonic operators. Far inside the Higgs phase 
	($\lambda=0.01,\kappa = 0.3900$) the string tension is vanishingly
	small. Mass ratios constructed using the lightest scalar mass
	(obtained from the purely scalar operator) as a reference do
	give good agreement for the operators of mixed origin,
	however.  Thus we have good evidence for scaling between
	$\beta=6.0$ and $9.0$, along lines of constant physics,
	and hence that our conclusions in this paper are relevant
	to the continuum theory.

        In conclusion, we have clarified the phase diagram
        of the $SU(2)$ adjoint Higgs model in 2+1 dimensions.
        For small values of the scalar self-coupling, $\lambda$,
        the expected first-order transition indeed occurs.
        But as we increase $\lambda$ this becomes a 
        crossover rather than a second-order transition.
        At the point where the first-order transition
        vanishes there appears to be a critical point where
        the scalar mass vanishes. These calculations
        have been performed for values of the lattice spacing 
        where deviations from the continuum limit are very small
        and so we believe that these are properties of the
        continuum theory. All this is, of course,
        very much reminiscent of the behaviour of the 
        theory with fundamental Higgs
\cite{fund_phase_diag}. 
 
        Understanding the phase diagram is an essential first
        step if we wish to use this theory as a testing ground
        for the monopole mechanism for confinement. The fact that
        the Higgs phase and the symmetric, confining phase are 
        smoothly connected through a crossover region, 
        makes us optimistic about the possibility of learning
        how confinement interpolates between the former phase,
        where it is well understood, \cite{polyakov}, and the
        latter, where theoretical ideas exist, \cite{thooft2},
        but still remain speculative. This optimism is
        reinforced by the simple behaviour that we have found 
        for the flux tube and the string tension as we pass
        through this crossover region. Of particular interest
        in this context is the critical point; it may be that
        we can find here a reasonably dense plasma of t'Hooft-Polyakov
        monopoles. These questions are under investigation.

	\vskip 0.20in
	\noindent {\bf Acknowledgements}

	The work of J.D.S. was supported in part by the National
	Science Foundation under Grant No. NSF PHY 94-12556 and,
	during his stay in Oxford, by a PPARC Visiting Fellowship,
	GR/K95345, and a grant from the Astor Fund of the University.
	That of A.G.H. was supported in part by the United States
	Department of Energy grant DE-FG05-91ER40617 and that of
	M.T. by PPARC grants GR/K55752 and GR/K95338. This work was
	initiated and partly carried out when all the authors
	were in Oxford. They are grateful to Oxford Theoretical   
	Physics for its hospitality and for providing most of the 
	computing resources for this work.


\begin{thebibliography}{99}


\bibitem{nadkarni} 
S.\ Nadkarni, 
{\it Nucl. Phys. B334 (1990) 559.}

\bibitem{MS} 
K.\ Farakos, K.\ Kajantie, K.\ Rummukainen and M.\ Shaposhnikov,
{\it Nucl. Phys. B425 (1994) 67;} 
A.\ Jakov\'{a}c, K.\ Kajantie and A.\ Patk\'{o}s, 
{\it Phys. Rev. D49 (1994) 6810.}

\bibitem{laine} 
M.\ Laine, 
{\it Nucl. Phys. B451 (1995) 484.}
See the results in Appendix A.1.

\bibitem{fundamental} 
O.\ Philipsen, M.\ Teper and H.\ Wittig,
{\it Nucl. Phys. B469 (1996) 445.}

\bibitem{bunk}  
B.\ Bunk, 
{\it Nucl. Phys. B (Proc. Suppl.) 42 (1995) 566.}

\bibitem{teper_3} 
M.\ Teper, 
{\it Phys. Lett. B289 (1992) 115.}

\bibitem{barb}
M.~N.~Barber, {\it in `Phase Transitions and Critical Phenomena', Vol.~8,}
eds.~C.Domb and J.~L.~Lebowitz, Academic Press, New York, 1983.

\bibitem{polyakov} A.\ M.\ Polyakov, 
{\it Nucl. Phys. B120 (1977) 429.}

\bibitem{thooft} G.\ 't Hooft, 
{\it Nucl. Phys. B79 (1974) 276.}

\bibitem{degrand}T.\ A.\ DeGrand and D.\ Toussaint,
{\it Phys. Rev. D 22 (1980) 2478.}

\bibitem{thooft2} G.\ 't Hooft, 
{\it Nucl. Phys. B190 (1981) 455.}

\bibitem{fund_phase_diag}
K.\ Kajantie, M.\ Laine, K.\ Rummukainen and M.\ Shaposhnikov,
{\it Phys.\ Rev.\ Lett.\ 77 (1996) 2887;}
F.\ Karsch, T.\ Neuhaus, A.\ Patk\'os, J.\ Rank, 
{\it Preprint FSU-SCRI-96C-79, hep-lat/9608087, 
to be published in the proceedings of Lattice 96.}

\end{thebibliography}
\end{document}